\documentclass[12pt]{article}
\usepackage{subeqn}
\usepackage{graphicx}
\usepackage{ifpdf}
\ifpdf  \fi
\usepackage{amsfonts,amssymb}
\usepackage{mathptmx,helvet,courier,makeidx,multicol,footmisc}
\usepackage[numbers]{natbib}
\bibpunct{(}{)}{;}{a}{,}{,}
\usepackage[bookmarksnumbered=true, pdfauthor={Wen-Long Jin}]{hyperref}

\newcommand{\commentout}[1]{}

\newcommand{\ba}{\begin{array}}
        \newcommand{\ea}{\end{array}}
\newcommand{\bc}{\begin{center}}
        \newcommand{\ec}{\end{center}}
\newcommand{\bdm}{\begin{displaymath}}
        \newcommand{\edm}{\end{displaymath}}
\newcommand{\bds} {\begin{description}}
        \newcommand{\eds} {\end{description}}
\newcommand{\ben}{\begin{enumerate}}
        \newcommand{\een}{\end{enumerate}}
\newcommand{\beq}{\begin{equation}}
        \newcommand{\eeq}{\end{equation}}
\newcommand{\bfg} {\begin{figure}[h]}
        \newcommand{\efg} {\end{figure}}
\newcommand{\bi} {\begin {itemize}}
        \newcommand{\ei} {\end {itemize}}
\newcommand{\bqn}{\begin{eqnarray}}
        \newcommand{\eqn}{\end{eqnarray}}
\newcommand{\bqs}{\begin{eqnarray*}}
        \newcommand{\eqs}{\end{eqnarray*}}
\newcommand{\bsl} {\begin{slide}[8.8in,6.7in]}
        \newcommand{\esl} {\end{slide}}
\newcommand{\bsq}{\begin{subequations}}
        \newcommand{\esq}{\end{subequations}}       
\newcommand{\bss} {\begin{slide*}[9.3in,6.7in]}
        \newcommand{\ess} {\end{slide*}}
\newcommand{\btb} {\begin {table}}
        \newcommand{\etb} {\end {table}}
\newcommand{\bvb}{\begin{verbatim}}
        \newcommand{\evb}{\end{verbatim}}

\newcommand{\m}{\mbox}
\newcommand {\der}[2] {{\frac {\m {d} {#1}} {\m{d} {#2}}}}

\newcommand {\pd}[2] {{\frac {\partial {#1}} {\partial {#2}}}}

\newcommand{\reff}[1] {{{Figure \ref {#1}}}}
\newcommand{\refe}[1] {{(\ref {#1})}}

\def\la      {{\lambda}}

\def\pmb#1{\setbox0=\hbox{$#1$}%
   \kern-.025em\copy0\kern-\wd0
   \kern.05em\copy0\kern-\wd0
   \kern-.025em\raise.0433em\box0 }

\def\eop{{\hfill $\blacksquare$}}

\newtheorem{theorem}{Theorem}[section]

\newtheorem{lemma}[theorem]{Lemma}

\def\dt     {{\Delta t}}

\usepackage{multirow}
\usepackage{subcaption}
\usepackage{float}

\begin{document}
	\title{Stable dynamic pricing scheme independent of lane-choice models for high-occupancy-toll lanes} 
\author{Wen-Long Jin \footnote{Department of Civil and Environmental Engineering, California Institute for Telecommunications and Information Technology, Institute of Transportation Studies, 4000 Anteater Instruction and Research Bldg, University of California, Irvine, CA 92697-3600. Tel: 949-824-1672. Fax: 949-824-8385. Email: wjin@uci.edu. Corresponding author} and Xuting Wang \footnote{Department of Civil and Environmental Engineering, Institute of Transportation Studies, University of California, Irvine} and Yingyan Lou \footnote{Department of Civil, Environmental and Sustainable Engineering, Arizona State University}}
\maketitle
\begin{abstract}
A stable dynamic pricing scheme is essential to guarantee the desired performance of high-occupancy-toll (HOT) lanes, where single-occupancy vehicles (SOVs) can pay a price to use the HOT lanes. But existing methods apply to either only one type of lane-choice models with unknown parameters or different types of lane-choice models but with known parameters. In this study we present a new dynamic pricing scheme that is stable and applies to different types of lane-choice models with unknown parameters.
 
There are two operational objectives for operating HOT lanes: (i) to maintain the free-flow condition to guarantee the travel time reliability; and (ii) to maximize the HOT lanes' throughput to minimize the system's total delay. The traffic dynamics on both HOT and general purpose (GP) lanes are described by point queue models, where the queueing times are determined by the demands and capacities. We consider three types of lane-choice models: the multinomial logit model when SOVs share the same  value of time, the vehicle-based user equilibrium model when SOVs' values of time are heterogeneous and follow a distribution, and a general lane-choice model. We demonstrate that the second objective is approximately equivalent to the social welfare optimization principle for the logit model. Observing that the dynamic price and the excess queueing time on the GP lanes are linearly correlated in all the lane-choice models, we propose a feedback control method to determine the dynamic prices based on two integral controllers. We further present a method to estimate the parameters of a lane-choice model once its type is known. Analytically we prove that the equilibrium state of the closed-loop system with constant demand patterns is ideal, since the two objectives are achieved in it, and that it is asymptotically stable. With numerical examples we verify the effectiveness of the solution method. 

\end{abstract}
 
\section{Introduction}
High-occupancy-toll (HOT) lanes have been one of the most successful lane management methods, which combine high-occupancy vehicle (HOV) lanes and congestion pricing strategies by charging single-occupancy vehicles (SOVs) to use HOV lanes during peak periods. At a congested corridor where the HOV lanes are under-utilized, they offer numerous benefits to both the operators and users by making use of the unutilized capacity of the HOV lanes. HOT lanes have been implemented on State Route 91, US Route 101, and I-15 in California and other places \citep{perez2003guide}. 

There are two operational objectives for operating HOT lanes: (i) maintaining the free-flow condition; and (ii) maximizing the flow rate of the HOT lanes. This will help to guarantee the trip time reliability of both HOVs and paying SOVs, and minimal congestion level on the general purpose (GP) lanes. 
From a control theoretic point of view, the actuation signal is the price charged to SOVs, and the system's response is reflected by the percentage of SOVs choosing the HOT lanes. Therefore, a core problem is to determine the dynamic prices for time-dependent and stochastic demands of both HOVs and SOVs, such that the two objectives are achieved. 

Intuitively, at the same price SOVs with higher values of time (VOTs) are more willing to pay and switch to HOT lanes. Consequently, with higher prices, fewer SOVs would pay and switch, given that all necessary information, including both saved time and price, is provided.\footnote{Note that drivers may consider the price as an indication of time savings and therefore are more willing to switch to the HOT lanes at higher prices \citep{janson2014hot}. Thus, timely and accurate information is critical to mitigate such psychological effects of pricing and help drivers make rational decisions.} Thus, SOVs' willingnesses to pay and VOTs are as important as prices in determining the number of SOVs choosing the HOT lanes over the GP lanes, which in turn determines the effectiveness of the HOT lanes. However, the SOVs'  willingnesses to pay and VOTs are generally unknown to the operator. To address this issue, system operators either have to estimate SOVs' willingnesses to pay and VOTs and then calculate the dynamic prices accordingly, or to determine dynamic prices independent of such lane choice behaviors.
In addition, the stochasticity and dynamics in the demand pattern and other features can drive the state away from the ideal one when two aforementioned objectives are achieved. Therefore, another challenge is to make sure that the dynamic pricing scheme stabilizes the system and prevents the system from drifting away from the ideal state.

In \citep{yin2009dynamic,lou2011optimal}, the HOT lane management problem was studied with different traffic flow models for a single bottleneck and a corridor respectively, but they share the same multinomial logit model for SOVs' lane choice behavior. The solution strategies in the two studies are also the same: the VOT was first estimated dynamically with a Kalman filter, and the dynamic price was then calculated from the multinomial logit model. 
In \citep{wang2017new,wang2020control}, the HOT lane management problem for a single bottleneck was studied with the point queue traffic flow model and the multinomial logit model as in \citep{yin2009dynamic}, but a simpler feedback control method was used to first estimate the VOT and then calculate the dynamic prices. Further it was shown, both analytically and numerically, that the closed-loop system is stable and converges to the ideal state. 

It is well known, however, that the applicability of the multinomial logit model depends on the IIA (independence of irrelevant alternatives) assumption, which is likely violated in this traffic system due to the correlation between a user's utilities of the two choice alternatives.
In addition, it is important to consider the effects of heterogeneity when analyzing the effects of transport policies, because ignoring heterogeneity may bias the estimated welfare effects of tolling \citep{arnott1988schedule}.  Although the normal distribution is commonly applied to model VOT heterogeneity, its unboundedness property implies unrealistic choice behaviors \citep{liu2007estimation}. In theoretical studies, the uniform distribution was used when discussing the welfare effects of first-best and second-best congestion pricing in a bottleneck model with heterogeneous drivers \citep{van2011congestion}. In empirical studies,  the lognormal distribution of VOTs was observed when analyzing the results of revealed preference surveys \citep{bhat2000incorporating, small2005uncovering, borjesson2014experiences}. However, the lognormal distribution can lead to overestimation of VOTs in upper percentiles; thus, \citep{hess2005estimation} suggested to use triangular and Johnson's distributions.

In \citep{gardner2013development}, the dynamic pricing problem was studied with three different lane-choice models: all-or-nothing, logit, and a vehicle-based user equilibrium model with a distribution of VOTs (a simplified variant of the Burr distribution).
With simulations, the median VOT was set to be $15\$/h$, and different values for the shape parameters were applied to evaluate the models.
 But the  VOT and the distribution of VOTs were assumed to be known to the system operator, and the dynamic prices are calculated from this information and congestion levels. Even though offline surveys can be used to determine such a distribution, its accuracy can vary with different combinations of users at different times, and online estimations of VOTs can potentially lead to more efficient pricing schemes. That is, if the lane-choice models are incorrectly used, the resulting dynamic prices may not be able to drive the system to the ideal state. In addition, such open-loop pricing strategies cannot compensate the impacts of various disturbances.

In this study we attempt to fill the gap in the literature and present a new method to solve the dynamic control and estimation problems of the same traffic system as in \citep{yin2009dynamic,wang2020control}. But as in \citep{gardner2013development}, we consider different lane-choice models of SOVs, including the multinomial logit model in \citep{yin2009dynamic,wang2020control}. However, the solution method is different from those in both groups of studies: (1) different from \citep{yin2009dynamic,lou2011optimal,wang2017new,wang2020control}, we first solve the control problem and then the estimation problem; to achive this, the controller is independent of the underlying lane-choice models and their parameters. However, if the underlying lane-choice model is given to the operator, the  VOT and the distribution of VOTs  can then be estimated; (2) different from \citep{gardner2013development}, our dynamic prices are determined by a feedback controller, which is applicable when the lane-choice models and their parameters are unknown to the operator. In addition, we analytically prove that the new dynamic pricing scheme is stable for different lane-choice models. The stability is essential to guarantee the desired performance of high-occupancy-toll (HOT) lanes, since, with unstable pricing schemes, the traffic system easily drifts away from the ideal state when disturbed by randomness or measurement errors. 

 The rest of the paper is organized as follows. In Section 2, we define variables and describe the traffic flow model. In Section 3, we present three different lane-choice models. In Section 4, we define the control and estimation problems, discuss the design principle of dynamic pricing, and present a new solution method. In Section 5, we analyze the equilibrium state of the closed-loop system and its stability property. In Section 6, we present the discrete-time models and numerical solutions. In Section 7, we conclude the study with discussions on future research topics.

\section{Definitions of variables and traffic flow model}

\subsection{Definitions of variables}

We consider a freeway corridor with two types of lanes: the HOT and GP lanes. We ignore the impacts of multiple on- and off-ramps and assume that there is only one origin and destination pair and one bottleneck on each type of lanes.
Thus we can  apply the point queue model \citep{Vickrey1969congestion,jin2015pqm} to model traffic dynamics on the two types of lanes. We define the following variables, as shown in \reff{ill_HOT}:
\bi
\item $q_1(t)$ and $q_2(t)$ are the time-dependent arrival rates (or demands) of HOVs and SOVs, respectively, at time $t$.
\item $C_1$ and $C_2$ are the bottleneck capacities of the HOT and GP lanes,  respectively.
\item $\la_1(t)$ and $\la_2(t)$ are the queue lengths on the HOT and GP lanes, respectively.
\item $w_1(t)$ and $w_2(t)$ are vehicles' queueing times at the bottlenecks on the HOT and GP lanes, respectively. $w(t)=w_2(t)-w_1(t)$ is the difference. Here we assume that vehicles have the same free-flow travel time on both types of lanes, which are thus omitted.
\ei

First, we assume that the system is congested during the study period $t\in[0,T]$; i.e.,
\bqs
\int_0^t (q_1(s)+q_2(s)) ds>(C_1+C_2)t.
\eqs
Second, we assume that   the demand (arrival rate) of HOVs is below the HOV lanes' capacity; i.e.,  $q_1(t) < C_1$.
Thus, the capacity of the HOv lanes is under-utilized, and single-occupancy vehicles (SOVs) can pay a price to use the HOT lanes. Third, we assume that HOVs can use the HOT lanes for free.

We further define the following variables: 
\bi
\item  $u(t)$ is the time-dependent price paid by SOVs who switch to the HOT lanes from the GP lanes.
\item $q_3(t)$ is the flow-rate of paying SOVs, which are SOVs but use the HOT lanes.  
\item $p(t)$ is the proportion of paying SOVs, where
\bqn
p(t)&=&\frac{q_3(t)}{q_2(t)} \in [0,1]. \label{def:p}
\eqn

\ei

\bfg
\bc
\includegraphics[width=4in]{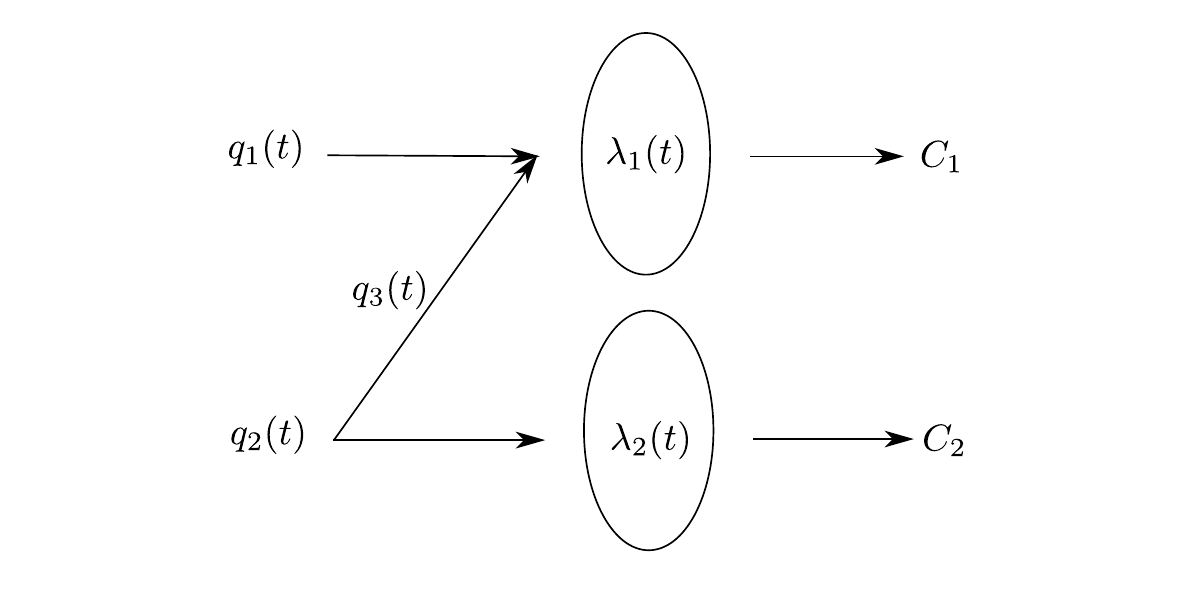}
\caption{Queueing representation of the traffic system with HOT and GP lanes}
\label{ill_HOT}
\ec
\efg

\subsection{Point queue models}
 For the point queues on the two lane groups, the arrival rates are respectively $q_1(t)+q_3(t)$ and $q_2(t)-q_3(t)$.
 Following \citep{wang2020control} we define the residual capacity of the HOT lanes as 
 \bqn
 \zeta(t)&=&C_1-q_1(t)-q_3(t). \label{def:zeta}
 \eqn
 Then the dynamics of the two point queues are described by the following ordinary differential equations \citep{jin2015pqm}:
 \bsq \label{dynamic_sys}
 \bqn
 \dot \la_1(t)&=&\max\bigg\{-\frac{\la_1(t)}{\epsilon}, -\zeta(t)\bigg\},\\
\dot \la_2(t)&=&\max\bigg\{-\frac{\la_2(t)}{\epsilon}, q_1(t)+q_2(t)-C_1-C_2+\zeta(t)\bigg\},
 \eqn
 \esq
 where the dot sign is for the derivative with respect to time, and $\epsilon=\lim_{\dt\to 0^+} \dt$ is an infinitesimal number  and equals $\dt$ in the discrete form. As shown in \citep{jin2015pqm}, such a formulation is well-defined, preventing negative queue lengths when queues dissipate.  
 
  The queueing times caused by the queues are given by $w_1(t)=\frac{\la_1(t)}{C_1}$, and $w_2(t)=\frac{\la_2(t)}{C_2}$. 
 The difference in the queueing times on the GP and HOT lanes is given by
 \bqn
 w(t)&=&\frac{\la_2(t)}{C_2}-\frac{\la_1(t)}{C_1}. \label{def:wt}
 \eqn
 
\section{lane-choice models}

Intuitively, the proportion of SOVs choosing the HOT lanes depends on the difference in the queueing times on the two types of lanes, $w(t)$, the price, $u(t)$, and the SOVs' social and economic characteristics. We represent the lane-choice model by the following functional relationship:
\bqn
p(t)&=&G(u(t),w(t)), \label{general-model-1}
\eqn
where $u(t),w(t)\geq 0$, and the function  $G(\cdot, \cdot)$ depends on the SOVs' characteristics, such as their VOTs.  
Here we assume that $p(t)$ does not depend on the historical values of $u(t)$ or $w(t)$, which would lead to differential or integral equations. We also assume that the drivers are fully aware of $u(t)$ and $w(t)$. In addition, we assume the existence of an underlying generalized cost or disutility function in terms of $u(t)$ and $w(t)$, such that some SOVs can reduce their costs by choosing the HOT lanes.

In addition, we assume that the lane-choice model, \refe{general-model-1}, satisfies the following conditions (with $u(t)\geq 0$ and $w(t)\geq 0$) with respect to the underlying lane-choice behaviors: 
\bsq\label{lane-choice-conditions}
\ben

\item The choice proportion decreases in $u(t)$; i.e., the higher the price, the fewer SOVs will choose to pay for the HOT lanes. Mathematically, if $G(u,w)$ is continuous and differentiable, we have
\bqn
\pd{G(u,w)}{u}<0.
\eqn
\item The choice proportion increases in $w(t)$; i.e., the larger the difference in queueing times, the more SOVs will choose to pay for the HOT lanes. Mathematically, we have
\bqn
\pd{G(u,w)}{w}>0.
\eqn 

\item Intuitively, the function $G(\cdot,\cdot)$ should also satisfy the following conditions:
 $G(u,\infty)=1$ for bounded $u\geq 0$, and $G(\infty,w)=0$ for bounded $w\geq 0$. That is, when the GP lanes are gridlocked and have infinite queueing times, then all SOVs would choose the HOT lanes; on the other hand, if the price is extremely high, no SOVs would choose the HOT lanes.

\een

\esq

From the respective definitions of $p(t)$ and $\zeta(t)$ in \refe{def:p} and \refe{def:zeta}, the above general lane-choice model leads to the following equation of $\zeta(t)$:
\bqn
\zeta(t)&=&C_1 -q_1(t)-q_2(t)\cdot G(u(t),w(t)), \label{general-model}
\eqn
which leads to the following relation between the residual capacity and the price:
\bqn
u(t)&=&\Phi(\zeta(t);w(t))\equiv G^{-1}(p(t);w(t))= G^{-1}\left(\frac{C_1-q_1(t)-\zeta(t)}{q_2(t)};w(t)\right). \label{resque-price}
\eqn
Note that the above function is an equivalent form of SOVs' lane-choice model and assumed to be unknown to the system operator. Therefore, the system operator cannot use \refe{resque-price} to directly calculate the price from the residual capacity, as in \citep{gardner2013development}, which assumes that the system operator is aware of the lane-choice model and its parameters.  From \refe{lane-choice-conditions} and \refe{general-model} we can see that
\bsq \label{lane-choice-conditions2}
\bqn
\pd{u}{\zeta}&=&-1/ [q_2(t) \pd{G(u,w)}u]>0,\\
\pd{u}{w}&=&-\pd{G(u,w)}w /\pd{G(u,w)}u>0.
\eqn
\esq
Note that $\pd{u}{\zeta}>0$ suggests that the price and the residual capacity are positively correlated. Thus if the residual capacity is positive and high, we need to reduce the price to reduce the residual capacity and increase the utilization of the HOT lanes. In contrast, if the residual capacity is negative and the HOT lanes are congested, we need to increase the price to increase the residual capacity. Similarly, $u(t)$ and $w(t)$ should also be positively correlated. These observations serve as the guidelines for designing the dynamic pricing scheme.

In the following, we present three examples of lane-choice models.

\subsection{Logit model}

If SOVs choose the HOT lanes based on the logit model, then \refe{general-model-1} can be written as 
\bqn
p(t)&=&1/[1+e^{\alpha_*[u(t)- \pi_* w(t)] }], \label{logit_proportion}
\eqn
where all SOVs are assumed to have the same VOT, $\pi_*$, $\alpha_*$ is the scale parameter, and the deterministic parts of the utilities on the HOT and GP lanes for SOVs are respectively $-\pi_*w_1(t)-u(t) $ and $-\pi_* w_2(t)$. Note that $\pi_*$ is unknown to the system operator. Correspondingly, \refe{resque-price} in this case can be written as
\bqn
u(t)&=&\pi_* w(t)+ \frac 1{\alpha_*}\ln \frac{q_1(t)+q_2(t)-C_1+\zeta(t)}{C_1-q_1(t)-\zeta(t)}. \label{u-logit}
\eqn

\subsection{Vehicle-based user equilibrium model with heterogeneous values of time}
We assume that SOVs have heterogeneous VOTs and denote the VOT of SOV $i$ by $\pi_i$, which varies with $i$.  We assume that the VOTs are identically and independently distributed random variables, which could be discrete or continuous or mixed. We denote the probability density function by $f(\pi)$, which is unknown to the system operator.

In the original Wardrop's user equilibrium (UE) state, ``the journey times on all the routes actually used are equal, and less than (or equal to) those which would be experienced by a single vehicle on any unused route''. 
Here we extend the UE principle for individual vehicles choosing different lanes based on generalized costs, such that ``the general cost on a lane actually used by an SOV is less than or equal to that on an unused lane''. We refer to this as vehicle-based UE principle. That is, if an SOV chooses a lane (HOT or GP), then the cost on the lane is less than or equal to that on the other lane (GP or HOT) which would be experienced by the same vehicle. The vehicle-based UE principle is consistent with the selfish or rational choice principle and can be easily applied with explicit VOTs and congestion pricing.

Therefore, if SOV $i$ chooses the HOT lanes at $t$, then
\bsq
\bqn
\pi_i w_1(t) +u(t) &\leq &\pi_i w_2(t) ; \label{hot_cond}
\eqn
if SOV $j$ chooses the GP lanes, then
\bqn
\pi_j  w_1(t) +u(t) &\geq & \pi_j w_2(t). \label{gp_cond}
\eqn
\esq
Without loss of generality, assuming that $\pi_i$ and $\pi_j$ are continuous  random variables and $w_2(t)>w_1(t)$,  we have the following conclusion.

\begin{lemma}
	At $t$, the proportion of SOVs choosing the HOT lanes is given by
	\bqn
	p(t)&=&1-F\left(\frac{u(t)}{w(t)}\right), \label{proportion_HOT}
	\eqn	
	where $F(\cdot)$ is the cumulative distribution function of $f(\cdot)$.
\end{lemma}
 {\em Proof}. From \refe{hot_cond}, we can see that, for any SOV $i$ choosing the HOT lanes, $\pi_i \geq \frac{u(t)}{w(t)}$;
for any SOV $j$ choosing the GP lanes, $\pi_j \leq \frac{u(t)}{w(t)}$.
Therefore, the proportion of SOVs choosing the HOT lanes is given by \refe{proportion_HOT}. \eop

Note that \refe{proportion_HOT} was first presented in \citep{gardner2013development}, but without relating it to the vehicle-based UE principle.

Correspondingly, \refe{resque-price} in this case can be written as
\bqn
u(t)&=&z \left( \frac{C_1-q_1(t)-\zeta(t)}{q_2(t)} \right) w(t),
\eqn
where $z(p)$ is $100(1-p)$th-percentile, defined by $p=1-F(z(p))$.
As an example, if the VOTs follow an exponential distribution, $F(x)=1-e^{-\frac{x}{\pi_*}}$, where $\pi_*$ is the  VOT, then we have
\bqn
u(t)&=& \pi_* w(t) \ln \frac{q_2(t)}{C_1-q_1(t)-\zeta(t)}  . \label{exp-distribution}
\eqn
As another example, if the VOTs follow the simplified variant of the Burr distribution in \citep{gardner2013development}, then we have
\bqn
u(t)&=& \pi_* w(t) \left(\frac{q_1(t)+q_2(t)-C_1+\zeta(t)}{C_1-q_1(t)-\zeta(t)}\right)^{1/\gamma}  ,
\eqn
where $\gamma$ is a shape parameter.

\subsection{A general lane-choice model}
In both logit and UE models of lane choices, the relationship between the price and the residual capacity can be written in the following form:
\bqn
u(t)&=&A(\zeta(t)) w(t)+B(\zeta(t)), \label{general-lc-model}
\eqn
which can be considered a general lane-choice model, an equivalent formulation of \refe{general-model}. From \refe{lane-choice-conditions2}, we can see that $\pd u w=A(\zeta)> 0$, and $\pd u \zeta=A'(\zeta) w+B'(\zeta)>0$ for any $w \geq 0$ and $\zeta$. Hence, 
\bqn
A'(\zeta)=\der{A(\zeta)}{\zeta}>0, \quad B'(\zeta)=\der{B(\zeta)}{\zeta}\geq0, \quad \m{and } A'(0)+B'(0)>0. \label{AB-properties}
\eqn
 Therefore, the price linearly increases in the excess queueing time on the GP lanes. Note that, however, $A(\zeta(t))$ and $B(\zeta(t))$ depend on the lane-choice models and their corresponding parameters, which again are both assumed to be unknown to the system operator.

\begin{figure}[H]
	\bc
	\includegraphics[scale=0.6]{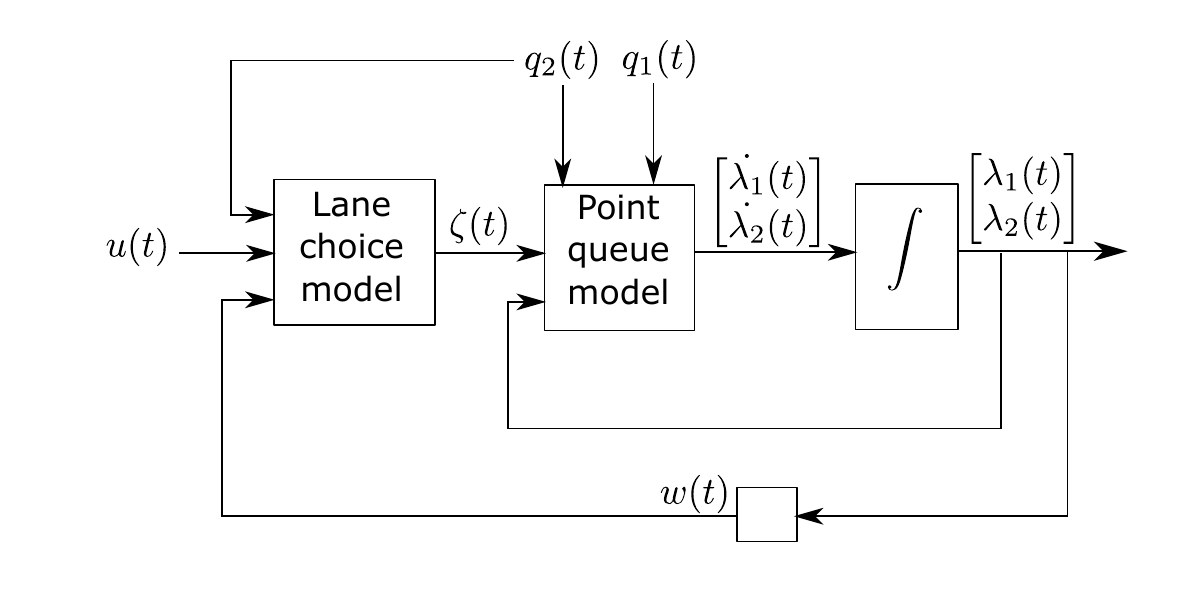}
	\caption{Block diagram of the traffic system}\label{plant}
	\ec
\end{figure}

Therefore, the point queue models for the queueing dynamics on both lanes in \refe{dynamic_sys}, the definition of the excess queueing time on the GP lanes in \refe{def:wt}, and the general lane-choice model in \refe{general-lc-model} describe the traffic system dynamics, as illustrated in \reff{plant}. In the traffic system, the inputs for the lane-choice model are $u(t)$, $q_2(t)$ and $w(t)$, and the output is $\zeta(t)$; for the traffic flow model (point queue model), the inputs are $q_1(t)$, $q_2(t)$ and $\zeta(t)$, and the outputs are $\la_1(t)$ and $\la_2(t)$.

\section{Control and estimation problems and their solutions}

For the traffic system with both GP and HOT lanes, there are two outstanding problems: in the control problem, we aim to determine the dynamic prices, $u(t)$, for time-dependent arrival rates of both HOVs and SOVs; and in the estimation problem, we aim to estimate the parameters in the lane-choice models, for example, the distribution of SOVs' VOTs, $f(\pi)$, for the vehicle-based user equilibrium model. 
In this study, we propose a new approach to solving the two problems: we first solve the control problem by finding the dynamic prices independent of the underlying lane-choice model and then the estimation problem by estimating the parameters of a given type of lane-choice models. This is different from the approaches in \citep{yin2009dynamic,lou2011optimal,wang2017new,wang2020control}, which first estimate the parameters for the multinomial logit model for lane choices and then calculate the dynamic prices. In addition, the approach in this study applies for different lane-choice models discussed in the preceding section; but those in the aforementioned references only apply for the multinomial logit model.  Since the dynamic prices are calculated without knowing the lane-choice model or its parameters in this study, this is also different from the method in \citep{gardner2013development}, where the dynamic prices are directly calculated from given distribution of VOTs or true VOT. 

 Here the traffic flow and lane-choice models are used to describe the dynamics of the traffic system, but not used in the design of the controller. That is, the dynamic pricing scheme is independent of both the traffic flow and lane-choice models. Thus the controller can potentially apply to other types of mathematical traffic flow models, simulators, or real-world systems.

\subsection{Two design principles for dynamic pricing schemes}
Generally, there are two operational objectives for the control problem \citep{perez2003guide}:  (i) to avoid congestion on the HOT lanes, and (ii) to maximize the HOT lanes' throughput and, therefore\footnote{As stated in Section 2.1, we assume that (i) the system is congested  and (ii) the demand of HOVs is below the HOT lanes’ capacity during the study period. Therefore, if the HOT lanes are fully utilized, the whole system is.}, the total throughput. They lead to two reference points for the system.
\ben
\item From the first operational objective, the reference point for the queue size on the HOT lanes is zero: $\la_1(t)=0$.
\item 
From the second operational objective, the optimal flow rate of SOVs choosing the HOT lane is $q_3(t)=C_1-q_1(t)$;
equivalently, the reference point for the residual capacity is zero: $\zeta(t)=0$.
\een

Naturally, the two objectives can serve as  two design principles for dynamic pricing schemes. The first principle is straightforward, since it is the hard constraint to avoid congestion on the HOT lanes. Next we discuss the relationship between the second principle and the social welfare maximization or total cost minimization principle.

When $\la_1(t)=0$, the HOVs are not impacted by the pricing scheme, and their costs are ignored in the total cost. For the paying SOVs, their waiting cost is zero as there is no queue on the HOT lanes, and the total cost equals the total prices, which during the study period $[0,T]$ can be written as 
\bqn
\phi_3&=& \int_0^T q_3(t) u(t) dt.
\eqn
For the SOVs staying on the GP lanes, the total cost is their queueing costs, which can be written as
\bqn
\phi_2&=&\int_0^T (q_2(t)-q_3(t)) w_2 (t) \bar \pi(q_2(t)-q_3(t)) dt,
\eqn
where $\bar \pi(q_2(t)-q_3(t))$ is the  VOT of these SOVs. Thus, the optimal $q_3(t)$ should solve the following constrained optimization problem:
\bsq \label{costminimization}
\bqn
\min_{q_3(t)} \phi\equiv \phi_2+\phi_3,
\eqn
s.t.
\bqn
q_3(t)&\leq &C_1-q_1(t),\\
q_3(t)&\geq&0,
\eqn
as well as the traffic flow and lane-choice models.
\esq

\begin{theorem} \label{principle_equivalence}
	For the logit model of lane choice, the cost minimization problem \refe{costminimization} is approximately solved by $q_3(t)=C_1-q_1(t)$ or $\zeta(t)=0$, when the demands are approximately constant.
\end{theorem}
{\em Proof}.
With the logit lane-choice model, we have from \refe{u-logit} 
\bqs
\phi_3&=&\int_0^T q_3(t) [\pi_* w_2(t) +\frac 1{\alpha_*} \ln \frac{q_2(t)-q_3(t)}{q_3(t)}] dt,
\eqs
and 
\bqs
\phi_2&=&\int_0^T (q_2(t)-q_3(t)) \pi_*  w_2 (t)  dt.
\eqs
Thus
\bqs
\phi&=&\int_0^T [\pi_* q_2(t)  w_2(t)  +  \frac 1{\alpha_*} q_3(t) \ln \frac{q_2(t)-q_3(t)}{q_3(t)}] dt.
\eqs
Since the GP lanes are always congested, and the queue length grows with time, we have
\bqs
w_2(t) \approx (q_2(t) -q_3(t)-C_2) t,
\eqs
when the demands are approximately constant.
Further since $\frac 1{\alpha_*} q_3(t) \ln \frac{q_2(t)-q_3(t)}{q_3(t)}$ is bounded, the total cost can be approximated by
\bqs
\phi&\approx& \int_0^T \pi_* q_2(t)(q_2(t) -q_3(t)-C_2) t dt,
\eqs
which decreases in $q_3(t)$. Therefore, the optimal $q_3(t)$ is its maximum value $C_1-q_1(t)$. \eop

Thus, Theorem \ref{principle_equivalence} establishes that the second principle approximates the social welfare optimization principle.

\subsection{A feedback method for solving the control problem}

Note that both $\zeta(t)$ and $\lambda_1(t)$ cannot be directly forced to be zero by the system operator. Instead, such ideal states have to be attained by setting appropriate prices, $u(t)$. In this subsection we present such a pricing scheme, based on the insights we have regarding the traffic flow and lane-choice models in the preceding sections. But the scheme is independent of the traffic flow and lane-choice models.

In the general lane-choice model in \refe{general-lc-model}, $A(\zeta)$ and $B(\zeta)$ are unknown to the system operator. But we can see that the price should be linearly increasing in the queueing time difference. Thus we propose the following controller
\bqn
u(t)&=&a(t) w(t)+b(t), \label{feedback-control}
\eqn
where $a(t)$ and $b(t)$ are determined by the following I-controllers (Integral-controllers) \citep[][Chapter 10]{astrom}:
\bsq\label{two-I-controllers}
\bqn
\dot a(t)&=&K_1 \la_1(t)-K_2 \zeta(t),\\
\dot b(t)&=&K_3 \la_1(t)-K_4 \zeta(t).
\eqn
\esq
Here the units of $a(t)$ and $b(t)$ are $\$/min$ and $\$$ respectively, and those of $K_1$, $K_2$, $K_3$, and $K_4$ are $\$/veh/min^2$, $\$/veh/min$, $\$/veh/min$, and $\$/veh$, respectively.

If there is a queue on the HOT lanes; i.e., if $\la_1(t)>0$, we increase the price so that fewer SOVs choose the HOT lanes. If $q_3(t)< C_1-q_1(t)$; or if $\zeta(t)>0$, we decrease the price so that more SOVs choose the HOT lanes. Therefore, all of the coefficients in \refe{two-I-controllers}, including $K_1$, $K_2$, $K_3$, and $K_4$, should be positive.

\refe{feedback-control} is a feedback control method, where $w(t)$, $\la_1(t)$, and $\zeta(t)$ need to be measured in real time and fed back to the controller to update the dynamic price. Note that the type of lane-choice models or their parameters as well as the traffic flow model and its parameters are not needed for the control scheme.

\begin{figure}[H]
	\bc
	\includegraphics[scale=0.6]{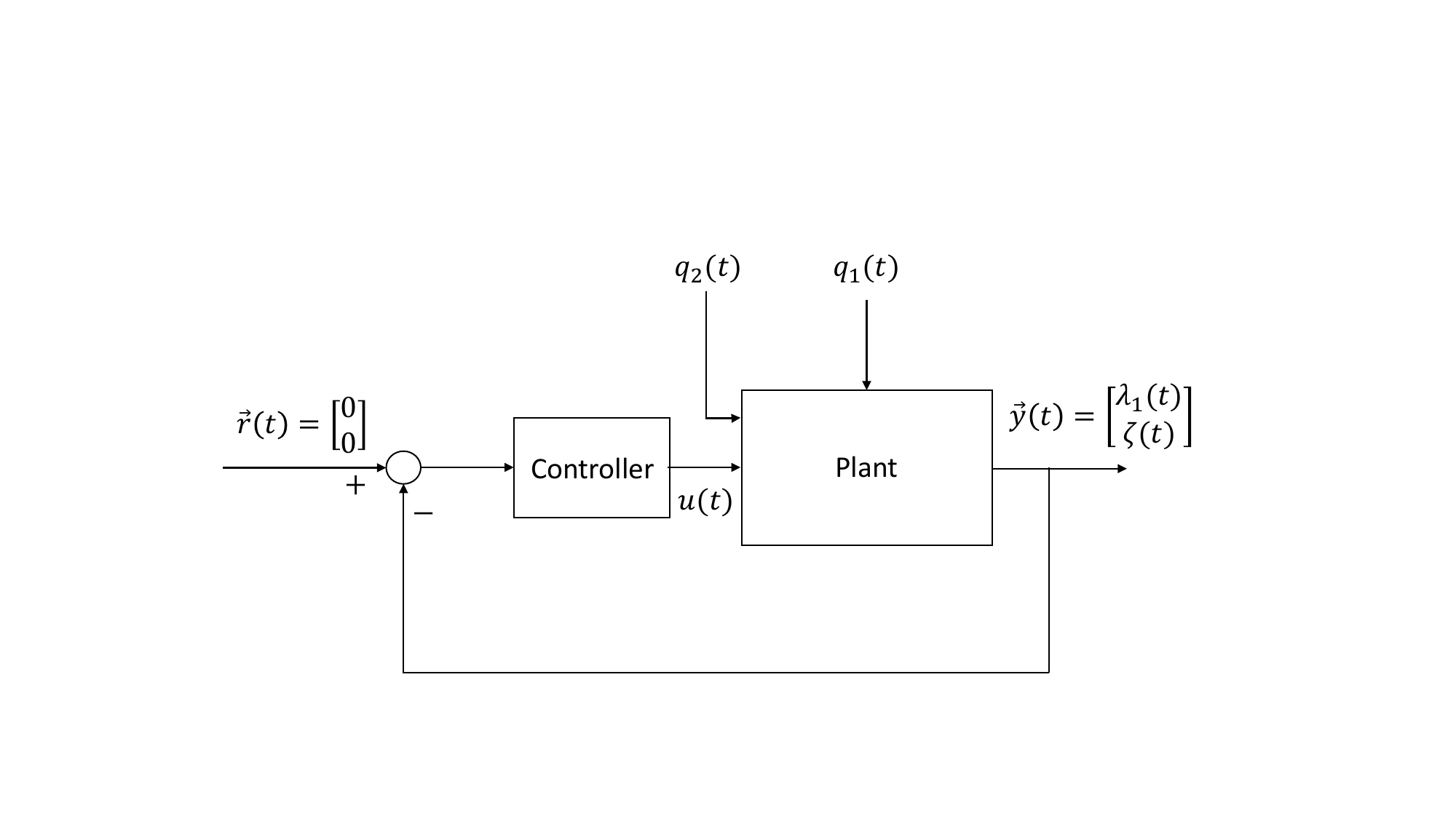}
	\caption{Block diagram of the closed-loop control system}\label{block}
	\ec
\end{figure}

The block diagram of the closed-loop control system is shown in \reff{block}. $q_1(t)$ and $q_2(t)$ are the inputs of the system; $\la_1(t)$ and $\la_2(t)$ are the state variables; $\vec y(t) =(\la_1(t)$,  $\zeta(t))$ are the outputs. Here $\vec r(t)$ is the reference signal for $\vec y (t)$. 
The traffic system (also known as the plant in control theory) consists of a lane-choice model and a traffic flow model, as shown in \reff{plant}. Dynamic prices are calculated by the controller. For the controller, the input signals are the differences between the reference signals and the outputs; i.e., $-\la_1(t)$ and $-\zeta(t)$, as illustrated by the $+$ and $-$ signs between $\vec r(t)$ and $\vec y(t)$.

\subsection{Estimation of values of time}
When designing the feedback controller in \refe{feedback-control}, the system operators assume no knowledge regarding drivers' VOTs. However, if the type of lane-choice models is known, we can estimate the corresponding parameters. In particular, we can estimate the VOT, $\pi_*$, for the logit model, and the probability density function of VOTs, $f(\pi)$, for the vehicle-based UE model.

If the lane choice behaviors are described by the logit model in \refe{logit_proportion},  we can estimate $\pi_*$ as
\bqn
\hat \pi_*(t)&=& \left[ u(t)-\ln \frac{q_2(t)-q_3(t)}{q_3(t)} \right]/w(t).
\eqn
That is, with the observed $q_2(t)$, $q_3(t)$, and $w(t)$ as well as the imposed price $u(t)$, we can obtain a time-dependent estimate of $\pi_*$, $\hat \pi_*(t)$. 

If the lane choice behaviors are described by the vehicle-based UE model in \refe{proportion_HOT}, we can estimate the cumulative distribution function of the VOTs as
\bqs
\hat F \left(\frac{u(t)}{w(t)}  \right)= 1-\frac{q_3(t)}{q_2(t)}.
\eqs
That is, with the observed $q_2(t)$, $q_3(t)$, and $w(t)$ as well as the imposed price $u(t)$, we can plot the cumulative distribution function $\hat F(\pi)=1-\frac{q_3(t)}{q_2(t)}$ against $\pi=\frac{u(t)}{w(t)}$. From the cumulative distribution function we can calculate the probability density function $\hat f(\pi)=\der{\hat F(\pi)}{\pi}$.

Note that $q_1(t)$ is not needed for estimating the VOTs. But $q_1(t)<C_1$ needs to be satisfied.
Such an estimation is more accurate with larger variations in $q_1(t)$ and $q_2(t)$. 
The estimation errors can be caused by the measurement errors as well as the errors in $u(t)$, which equal the differences between the optimal prices and the prescribed prices by the controller.

\section{Equilibrium state and  its stability property of the closed-loop system}
In this section we analyze the equilibrium state and its stability property of the closed-loop system, comprising of \refe{dynamic_sys}, \refe{def:wt}, \refe{general-lc-model}, \refe{feedback-control}, and \refe{two-I-controllers}. For simplicity, we assume that the demands are constant: $q_1(t)=q_1$, and $q_2(t)=q_2$.

\subsection{Equilibrium state}
In the equilibrium state, the queue size on the HOT lanes, $\la_1(t)$, $a(t)$, and $b(t)$ are all time-independent. That is, $\dot \la_1(t)=0$, $\dot a(t)=0$, and $\dot b(t)=0$. Then from (\ref{dynamic_sys}a) and \refe{two-I-controllers} we have
\bqs
\max\bigg\{ -\frac{\la_1(t)}\epsilon, -\zeta(t) \bigg\} &=&0,\\
K_1 \la_1(t)-K_2 \zeta(t)&=&0,\\
K_3 \la_1(t)-K_4 \zeta(t)&=&0,
\eqs
which are solved by $\la_1(t)=0$ and $\zeta(t)=0$. Therefore, the equilibrium state of the closed-loop system is the ideal state, where the two operational objectives are reached.

In the equilibrium state, (\ref{dynamic_sys}b) leads to $\dot \la_2(t)=C_2 w_0$, where 
\bqn
w_0&=&\frac{q_1+q_2-C_1-C_2}{C_2};
\eqn
thus,\footnote{Here we assume that the initial queue size on the GP lanes, $\la_2(0)=0$, without loss of generality.} $\la_2(t)=C_2 w_0 t$, and $w(t)=w_0 t$.
Both $a(t)=a$ and $b(t)=b$ are constant from \refe{two-I-controllers}, and the dynamic price in \refe{feedback-control} is given by 
\bqn
u(t)&=&a w_0 t+b. \label{equilibriumprice}
\eqn
Further from \refe{general-lc-model} we have $u(t)=A(0) w_0 t+ B(0)$.
Therefore we have $A(0)=a$, and $B(0)=b$. That is, $a(t)$ and $b(t)$ are the respective accurate estimators of $A(\zeta(t))$ and $B(\zeta(t))$ at the equilibrium state.

We can see that the dynamic price in the equilibrium state, \refe{equilibriumprice}, linearly increases in time. This result is different from directly applying the traditional I-controller to determine the dynamic price as in the following \citep{astrom,yin2009dynamic}: 
\bqn
\dot u(t)&=& K_1 \la_1(t)-K_2 \zeta(t),
\eqn
which would lead to a constant price in equilibrium state and fail to drive the closed-loop system to the ideal equilibrium state.

\subsection{Stability property of the equilibrium state}
From \refe{general-lc-model} and \refe{feedback-control} we can eliminate the control variable, $u(t)$:
\bqs
A(\zeta(t)) + \frac{B(\zeta(t))}{w(t)} &=& a(t) + \frac{b(t)}{w(t)}.
\eqs
Differentiating both sides with respect to $t$, we have
\bqs
\left[ A'(\zeta(t))+\frac{B'(\zeta(t))}{w(t)} \right] \dot \zeta(t)&=& \dot a(t)+\frac{\dot b(t)}{w(t)} +\frac{B(\zeta(t))-b(t)}{w^2(t)} \dot w(t).
\eqs
Substituting $\dot a(t)$ and $\dot b(t)$ in \refe{two-I-controllers} into the above equation, we obtain
\bqn
\left[ A'(\zeta(t))+\frac{B'(\zeta(t))}{w(t)} \right] \dot \zeta(t)&=& K_1 \la_1(t)-K_2 \zeta(t) \nonumber \\
&&+\frac{K_3 \la_1(t)-K_4 \zeta(t)}{w(t)} +\frac{B(\zeta(t))-b(t)}{w^2(t)} \dot w(t). \label{simplified-eqn}
\eqn

Here we consider the stability property of the equilibrium state when $t$ is sufficiently large. In this case, $w(t)$ can be approximated by $w_0 t$ near the equilibrium state. Also near the equilibrium state, $\zeta(t)$ is very small, and the left-hand side of \refe{simplified-eqn} can be approximated by $[A'(0)+\frac{B'(0)}{w_0 t}]\dot \zeta(t)$.   Since $K_3 \la_1(t)-K_4 \zeta(t)$, $B(\zeta(t))-b(t)$, and $\dot w(t)$ are all bounded near the equilibrium state, the right-hand side of \refe{simplified-eqn} can be approximated by $K_1 \la_1(t)-K_2 \zeta(t)$. Therefore, near the equilibrium state after a long time, the closed-loop system can be approximated by the following system of $(\la_1(t),\zeta(t))$:
\bsq\label{approximate-system}
\bqn
\dot \la_1(t)&=&\max\bigg\{-\frac{\la_1(t)}\epsilon, -\zeta(t)\bigg\},\\
\dot \zeta(t)&\approx&\frac1{A'(0)+\frac{B'(0)}{w_0 t}}\left( K_1 \la_1(t)-K_2 \zeta(t) \right).
\eqn
\esq
Note that \refe{approximate-system} is a switching system with two modes \citep{liberzon2003switching}:
\bi
\item When $\frac{\la_1(t)}\epsilon \leq \zeta(t)$; i.e., when the queue size on the HOT lanes is very small with $\la_1(t)\leq \epsilon \zeta(t)$, (\ref{approximate-system}a) leads to $\la_1(t)=0$ (the queue vanishes), and (\ref{approximate-system}b) can be simplified as
\bqn
\dot \zeta(t)&\approx&-\frac{K_2}{A'(0)+\frac{B'(0)}{w_0 t}} \zeta(t).\label{approximate-system-1}
\eqn
\item Otherwise, when the queue size on the HOT lanes is relatively large; i.e., when $\frac{\la_1(t)}\epsilon > \zeta(t)$, the queue persists, and \refe{approximate-system} can be simplified as
\bsq\label{approximate-system-2}
\bqn
\dot \la_1(t)&=& -\zeta(t),\\
\dot \zeta(t)&\approx&\frac1{A'(0)+\frac{B'(0)}{w_0 t}}(K_1 \la_1(t)-K_2 \zeta(t)).
\eqn
\esq
\ei
Both \refe{approximate-system-1} and \refe{approximate-system-2} are linear systems. Thus, the feedback controller in \refe{feedback-control} drives the traffic system dynamics to an approximate switching linear system after a long time. In spirit, the controller is a feedback linearization approach to nonlinear systems  \citep{khalil2002nonlinear}, but our approach is not a standard one.

\begin{theorem} \label{thm:stability}
With positive coefficients $K_1$ and $K_2$, the approximate switching linear system, \refe{approximate-system}, is asymptotically stable.
\end{theorem}
{\em Proof}. In the general lane-choice model \refe{general-lc-model},  $A'(0)+\frac{B'(0)}{w_0 t}>0$ from \refe{AB-properties}. Then it is straightforward that the eigenvalue of \refe{approximate-system-1} is negative, and the real parts of the two eigenvalues of \refe{approximate-system-2} are also negative. Thus both linear systems are asymptotically stable, and the corresponding switching system is asymptotically stable too.
\eop

\section{Numerical examples}

In the discrete model we set $\epsilon=\dt$, which is the time-step size. 
Given $\la_1(t)$ and $\la_2(t)$, $w(t)$ can be calculated from \refe{def:wt}. Given $a(t)$ and $b(t)$, we can calculate $u(t)$ from \refe{feedback-control}. Further from $q_1(t)$, $q_2(t)$, and $G(u(t),w(t))$, we can then calculate the residual capacity $\zeta(t)$ from \refe{general-model}. Further from the discrete versions of \refe{dynamic_sys} we can update the queue lengths at $t+\dt$:
\bsq
\bqn
\la_1(t+\dt)&=&\max\{0, \la_1(t)-\zeta(t) \dt\},\\
\la_2(t+\dt)&=&\max\{0, \la_2(t)+(q_1(t)+q_2(t)-C_1-C_2+\zeta(t))\dt\}.
\eqn
\esq
From the discrete versions of \refe{two-I-controllers}, $a(t+\dt)$ and $b(t+\dt)$ can be updated  as follows:
\bsq
\bqn
a(t+\dt)&=&a(t)+(K_1 \la_1(t)-K_2 \zeta(t))\dt,\\
b(t+\dt)&=&b(t)+(K_3 \la_1(t)-K_4 \zeta(t))\dt.
\eqn
\esq
In the discrete version of the closed-loop system, the following initial conditions are needed: $\la_1(0)$, $\la_2(0)$, $a(0)$, and $b(0)$; the following boundary conditions are needed: $q_1(t)$, and $q_2(t)$ at any time $t$; and the following parameters are needed: $K_1$, $K_2$, $K_3$, and $K_4$.
In this section, we set $K_1= 0.1$ \$/veh/$\mathrm{min}^{2}$, $K_2 = 0.1$ \$/veh/min, $K_3= 0.2$  \$/veh/min, and $K_4=0.2$ \$/veh; the initial values are $a(0) = 0.25$ \$/min and $b(0) = 0.1 \$ $.

The study site is a freeway corridor with one HOT and one GP lane, and the capacity for each lane is $30$ veh/min. The initial queue lengths on the HOT and GP lanes are 1 and 2, respectively ($\lambda_1(0)=1$ veh and $\lambda_2(0)=2$ veh). The simulation period is 20 minutes, and the time-step size is $\frac{0.1}{60}$ min. Note that $\dt$ should be sufficiently small; otherwise the computational errors could destabilize the system. For example, if we set $\Delta t = 1/60$ min, the system may not converge to the ideal equilibrium state after a long time (when the simulation time is longer than 250 minutes).

\subsection{The logit model}\label{simu-logit}
  When the SOVs follow the logit model for lane choices, we assume the true VOT is $0.5$ \$/min ($\pi_*= 0.5$ \$/min) and set the scale parameter $\alpha_*=1$. The demand of HOVs is constant at $q_1(t)=10$ veh/min, and the demand of SOVs is constant at $q_2(t)=60$ veh/min.
In \reff{logit-q}, the queue length on the HOT lane keeps increasing for the first minute, and the maximum queue length is about $2.8$ veh. After about 3 minutes, the queue length on the HOT lane drops to 0, and the queue length on the GP lane increases with time. When we increase $K_1$ and $K_2$ alone, the queue length on the HOT lane converges to 0 faster; and when we increase $K_3$ and $K_4$ alone, there is a smaller maximum queue length on the HOT lane. As shown in \reff{logit-u}, the price increases linearly after the queue on the HOT lane is eliminated. The price changing rate is about $0.2$ \$/min when the system reaches the ideal state. From \reff{logit-vot}, we find that the estimated VOT converges to the true value almost instantaneously. We conclude that the method is effective with the logit model. 

\bfg
\centering
\begin{subfigure}[c]{0.45\textwidth}
	\centering
	\includegraphics[width=\textwidth]{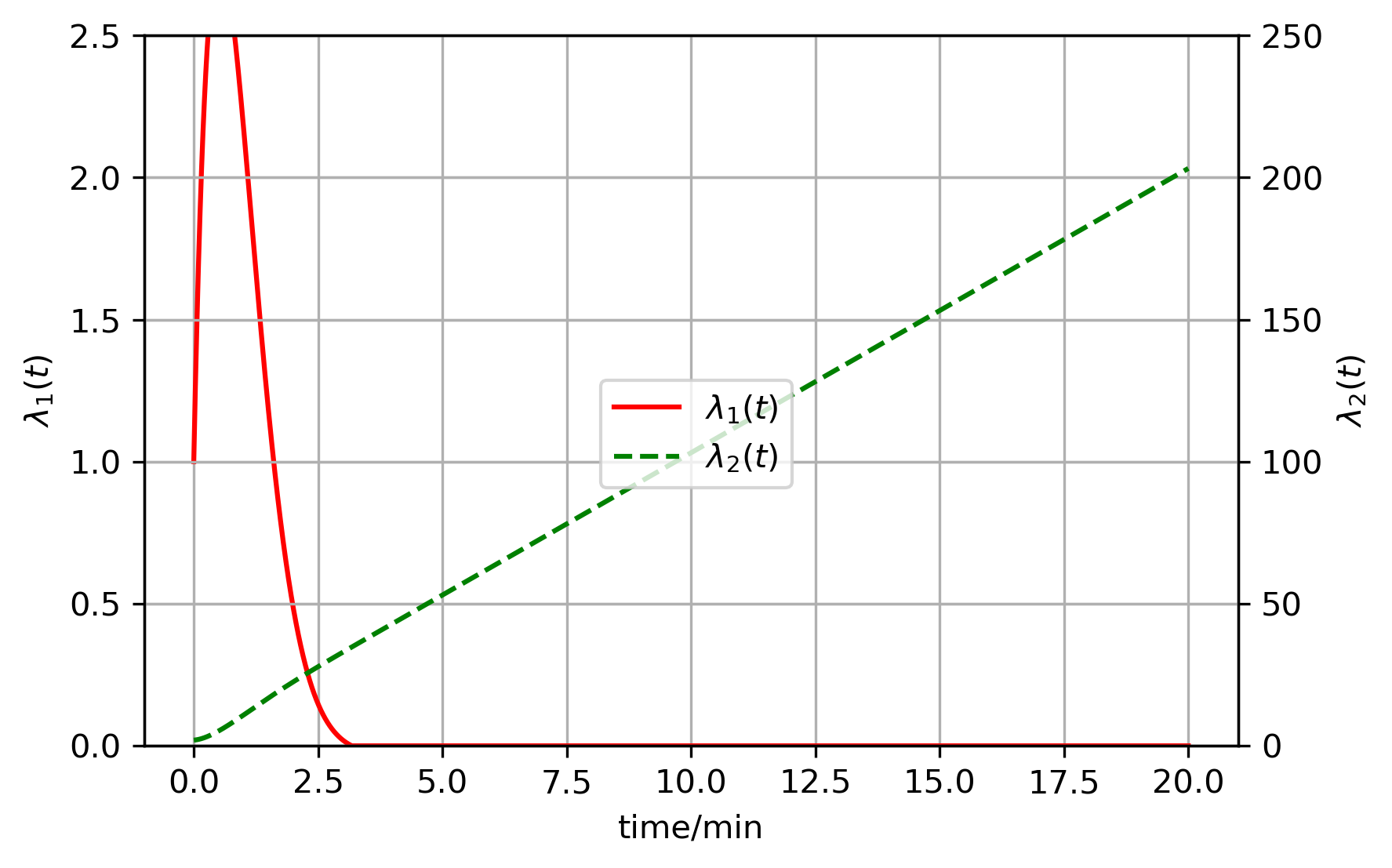}
	\caption[]%
	{{\small Queue lengths}}    
	\label{logit-q}
\end{subfigure}
\hfill
\begin{subfigure}[c]{0.45\textwidth}  
	\centering 
	\includegraphics[width=\textwidth]{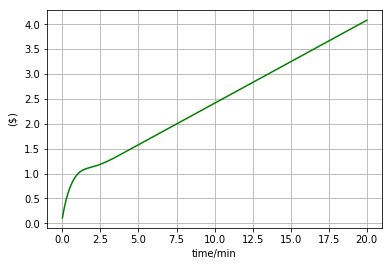}
	\caption[]%
	{{\small  Price for the HOT lanes}}    
	\label{logit-u}
\end{subfigure}
\vskip\baselineskip
\begin{subfigure}[c]{0.45\textwidth}   
	\centering 
	\includegraphics[width=\textwidth]{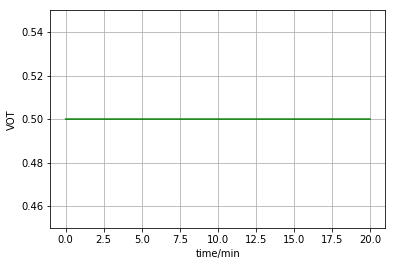}
	\caption[]%
	{{\small Estimated VOT}}    
	\label{logit-vot}
\end{subfigure}
\caption[]
{\small Simulation results with  the logit model and constant demands} 
\label{logit}
\efg

\bfg
\centering
\begin{subfigure}[c]{0.45\textwidth}
	\centering
	\includegraphics[width=\textwidth]{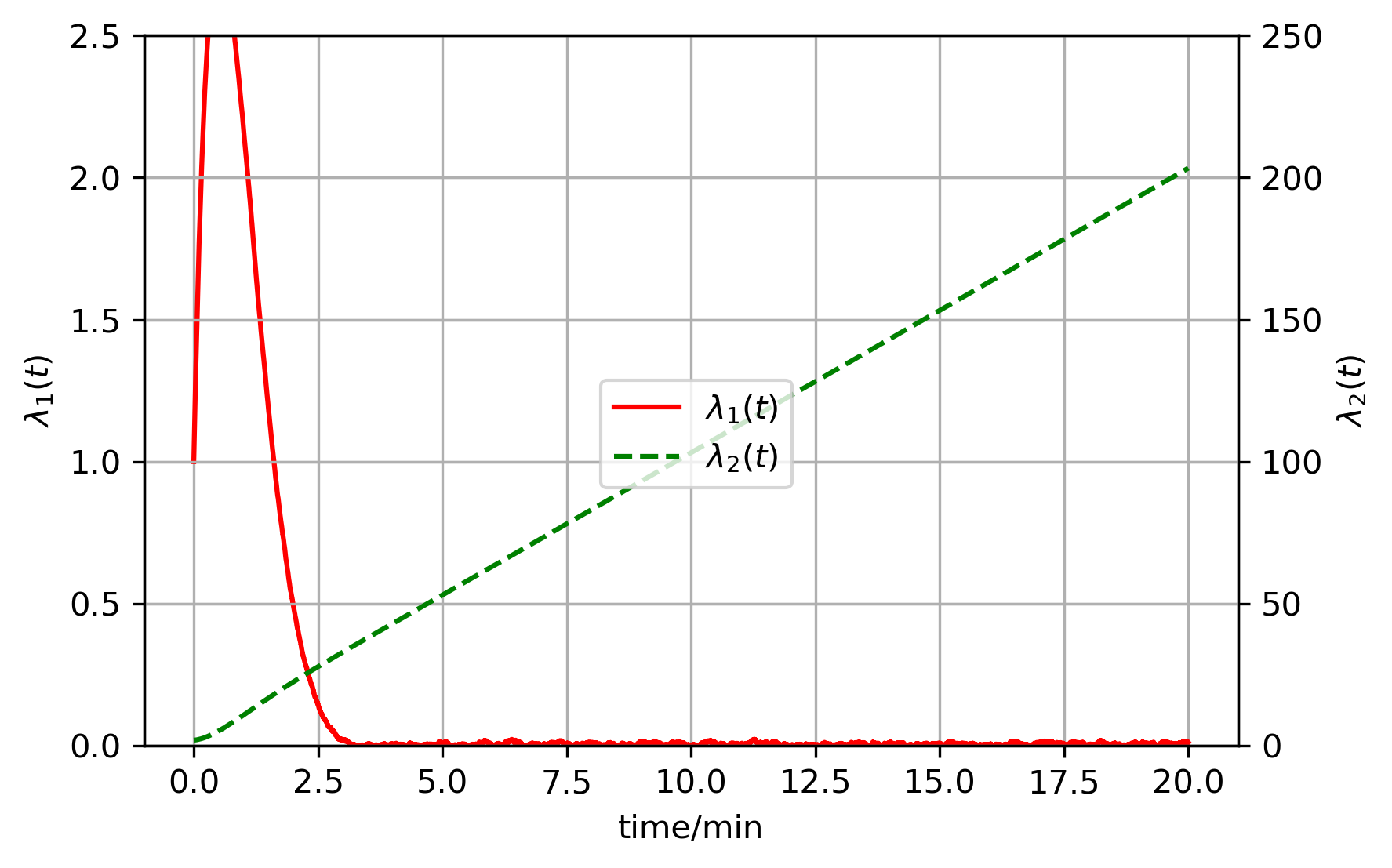}
	\caption[]%
	{{\small Queue lengths}}    
	\label{logit-q-random}
\end{subfigure}
\hfill
\begin{subfigure}[c]{0.45\textwidth}  
	\centering 
	\includegraphics[width=\textwidth]{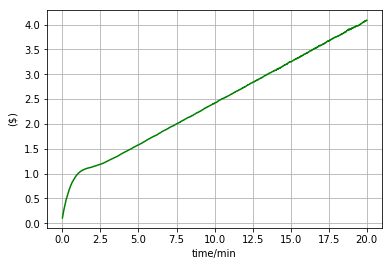}
	\caption[]%
	{{\small  Price for the HOT lanes}}    
	\label{logit-u-random}
\end{subfigure}
\vskip\baselineskip
\begin{subfigure}[c]{0.45\textwidth}   
	\centering 
	\includegraphics[width=\textwidth]{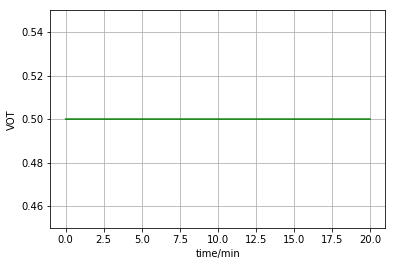}
	\caption[]%
	{{\small Estimated  VOT}}    
	\label{logit-vot-random}
\end{subfigure}
\caption[]
{\small Simulation results with  the logit model and random demands} 
\label{logit-random}
\efg

We next examine the robustness of the designed controller. In particular, we are interested in how the closed-loop system behaves when disturbances exist in the demand pattern. We assume $q_1(t)$ is a Poisson random variable with an average of 10 veh/min, and $q_2(t)$ is a Poisson random variable with an average of 60 veh/min. 
Similar to \reff{logit-q}, the queue length on the HOT lane converges to 0 after around 3 minutes (see \reff{logit-q-random}). Due to the randomness in the demand pattern, the queue length on the HOT lane is not always 0. However, it stays close to 0 eventually. In \reff{logit-u-random}, the dynamic price is similar to that in \reff{logit-u}: it increases almost linearly with time. The true  VOT can still be estimated with the disturbance (see \reff{logit-vot-random}). In this sense, we can conclude that the controller is robust with respect to random disturbances in the demand patterns.

\bfg
\begin{subfigure}{0.45\textwidth}
\centering
\includegraphics[width=\textwidth]{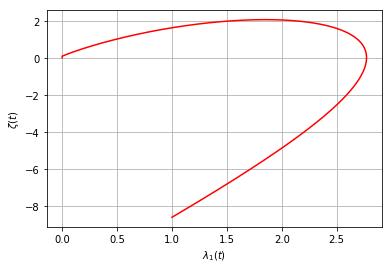}
\caption{}
\label{phase1}
\end{subfigure}
\hfill
\begin{subfigure}{0.45\textwidth}
\centering
\includegraphics[width=\textwidth]{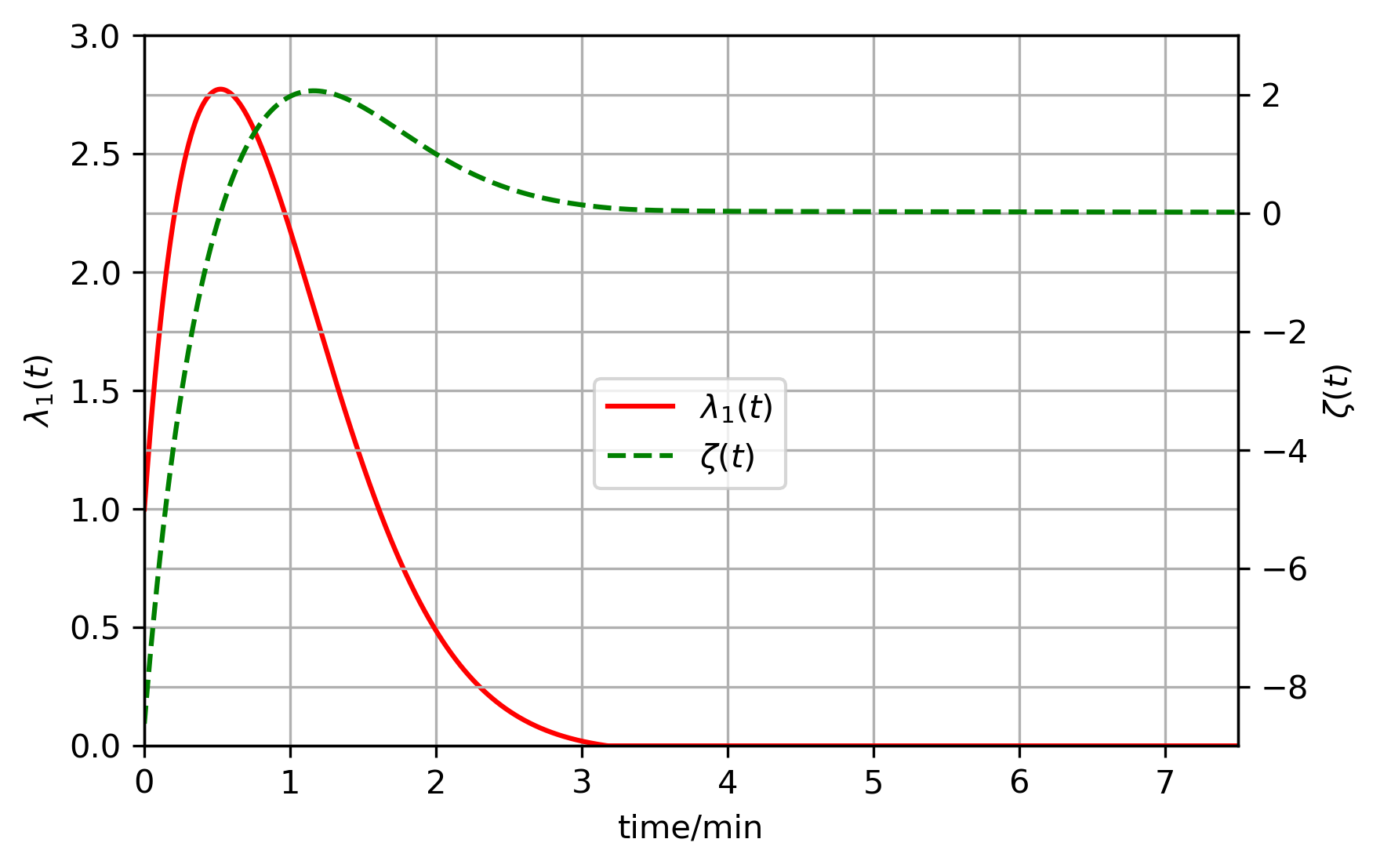}
\caption{}
\label{timeseries1}
\end{subfigure}
\caption[]
{\small Numerical verification of stability of the closed-loop system with the logit model}
\label{logit_original_mode1}
\efg

In the following, we numerically verify the stability property of the closed-loop system.
\reff{phase1} is the phase diagram: the horizontal axis represents $\lambda_1(t)$, and the vertical axis  $\zeta(t)$. The starting point is (1,-8.6) based on \refe{general-model}. Initially, $\zeta(t)$ increases till reaching 2.1  veh/min, and the maximum $\lambda_1(t)$ is 2.8 veh. It is clear that $\lambda_1(t)$ reaches 0 earlier than $\zeta(t)$. \reff{timeseries1} shows how $\lambda_1(t)$ and $\zeta(t)$ change with time: the horizontal axis is the simulation time, the left vertical axis represents $\lambda_1(t)$, and the right vertical axis represents $\zeta(t)$. $\lambda_1(t)$ reaches 0 at around 3 minutes, and after that $\zeta(t)$ converges to 0. With different parameters in the controller and different initial conditions, the system always converges to the ideal equilibrium state, as predicted by Theorem \ref{thm:stability}.

\subsection{The vehicle-based UE model}\label{simu-ue}
In this subsection, we assume that the VOTs follow an exponential distribution: $F(x)=1-e^{-2x}$ and SOVs follow the vehicle-based UE model for lane choices. 

\bfg
\centering
\begin{subfigure}[c]{0.45\textwidth}
	\centering
	\includegraphics[width=\textwidth]{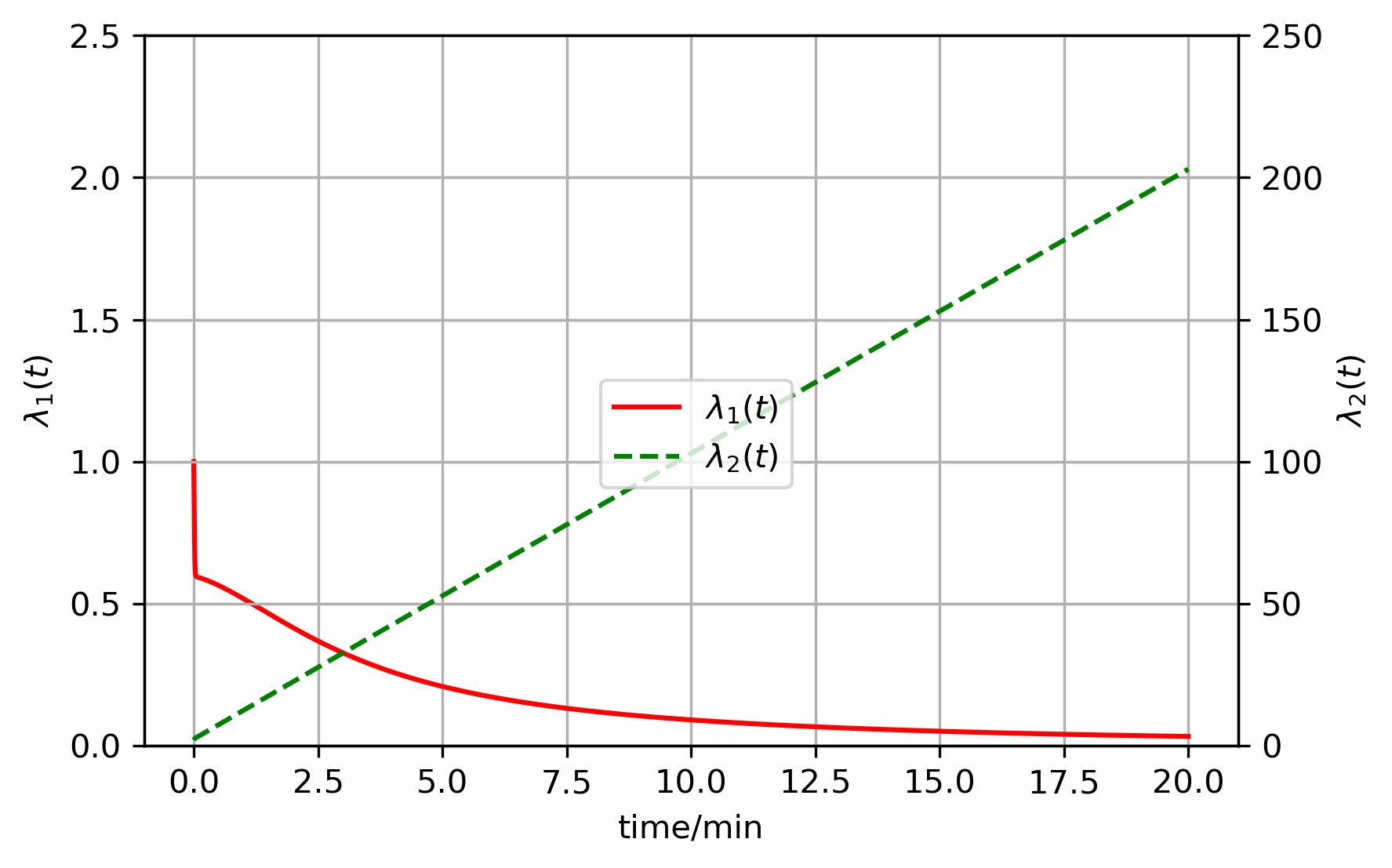}
	\caption[]%
	{{\small Queue lengths}}    
	\label{UE-q}
\end{subfigure}
\hfill
\begin{subfigure}[c]{0.45\textwidth}  
	\centering 
	\includegraphics[width=\textwidth]{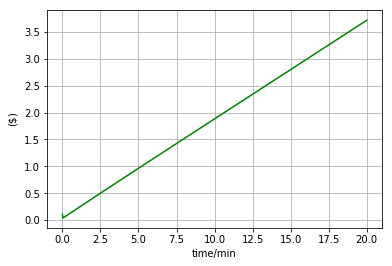}
	\caption[]%
	{{\small  Price for the HOT lanes}}    
	\label{UE-u}
\end{subfigure}
\caption[]
{\small Simulation results with  the vehicle-based UE model and constant demands} 
\label{ue}
\efg

For the same constant demands in Section \ref{simu-logit}, the initial price for the HOT lanes is $0.1\$$, which is too high for SOVs. In this case, almost no SOVs want to pay and switch to the HOT lane. Then, a sharp drop in the queue length on the HOT lanes occurs at the beginning, as shown in \reff{UE-q}.  The queue length on the HOT lane decreases to zero exponentially. The queue on the GP lane increases almost linearly with time. Different values for parameters in the controller are tested as well. When we increase $K_1$ and $K_2$ alone, the queue length on the HOT lane converges faster to the ideal state. In \reff{UE-u}, the price increases almost linearly in time with the rate of about 0.2 \$/min.

\bfg
\begin{subfigure}{0.45\textwidth}
	\centering
	\includegraphics[width=\textwidth]{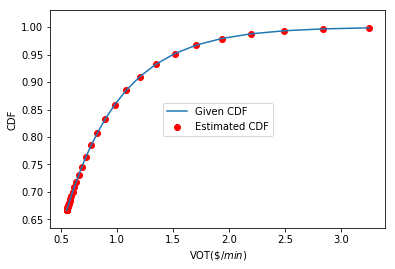}
	\caption{Estimated vs ground-truth CDF of VOTs}
	\label{cdf}
\end{subfigure}
\hfill
\begin{subfigure}{0.45\textwidth}
	\centering
	\includegraphics[width=\textwidth]{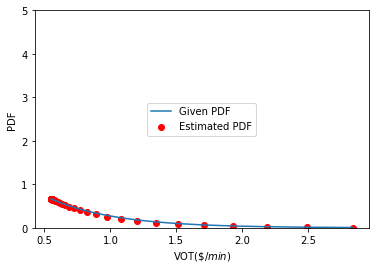}
	\caption{Estimated vs ground-truth  PDF of VOTs}
	\label{pdf}
\end{subfigure}
\caption{Estimation of the VOT distribution for the SOVs with the vehicle-based UE model}
\label{distribution}
\efg

\bfg
\begin{subfigure}{0.45\textwidth}
	\centering
	\includegraphics[width=\textwidth]{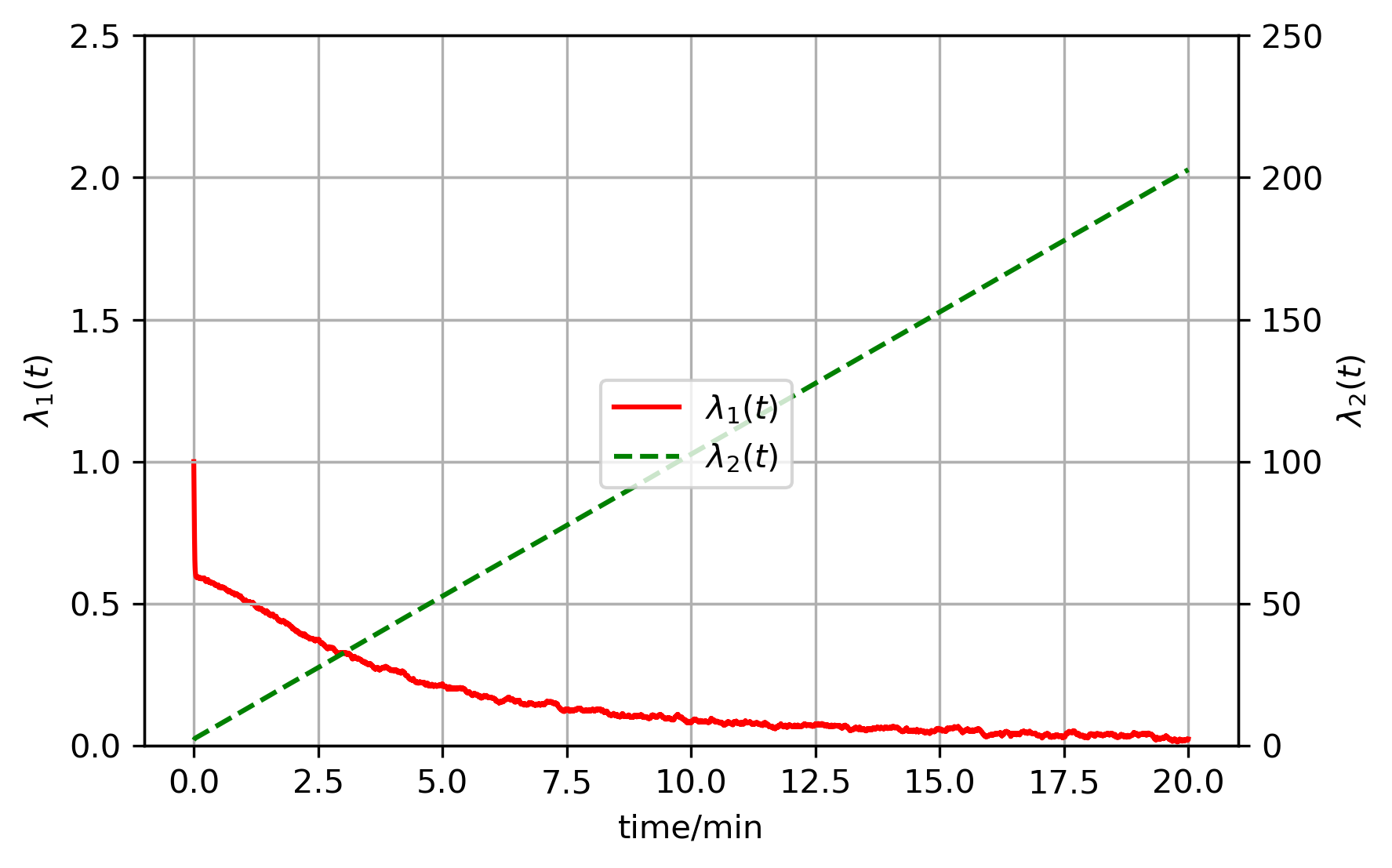}
	\caption{Queue lengths}
	\label{UE-random-q}
\end{subfigure}
\hfill
\begin{subfigure}{0.45\textwidth}
	\centering
	\includegraphics[width=\textwidth]{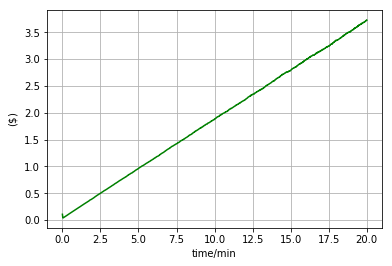}
	\caption{Price for the HOT lanes}
	\label{UE-random-u}
\end{subfigure}

\vskip\baselineskip
\begin{subfigure}{0.45\textwidth}
	\centering
	\includegraphics[width=\textwidth]{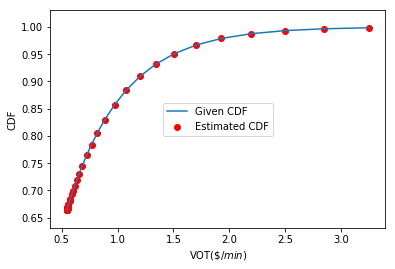}
	\caption{Estimated vs ground-truth CDF of VOTs}
	\label{UE_cdf_random}
\end{subfigure}
\hfill
\begin{subfigure}{0.45\textwidth}
	\centering
	\includegraphics[width=\textwidth]{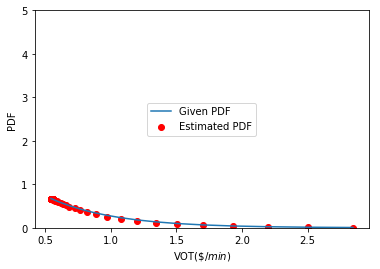}
	\caption{Estimated vs ground-truth  PDF of VOTs}
	\label{UE_pdf_random}
\end{subfigure}
\caption{Simulation results with the vehicle-based UE model and  random demands}
\label{UE_distribution_random}
\efg

Next we estimate the distribution of the VOTs. The cumulative distribution function (CDF) is  $F(x)=1-e^{-2x}$, so it is straightforward that the corresponding probability density function (PDF) is $f(x)=2e^{-2x}$. 
In \reff{distribution}, the red dots show the estimated CDF and PDF, and the solid lines represent the ground-truth functions. It is obvious that both estimated CDF and PDF are consistent with the ground-truth functions. 

\bfg
\begin{subfigure}{0.45\textwidth}
	\centering
	\includegraphics[width=\textwidth]{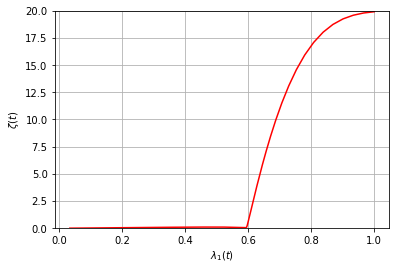}
	\caption{}
	\label{phase2}
\end{subfigure}
\hfill
\begin{subfigure}{0.45\textwidth}
	\centering
	\includegraphics[width=\textwidth]{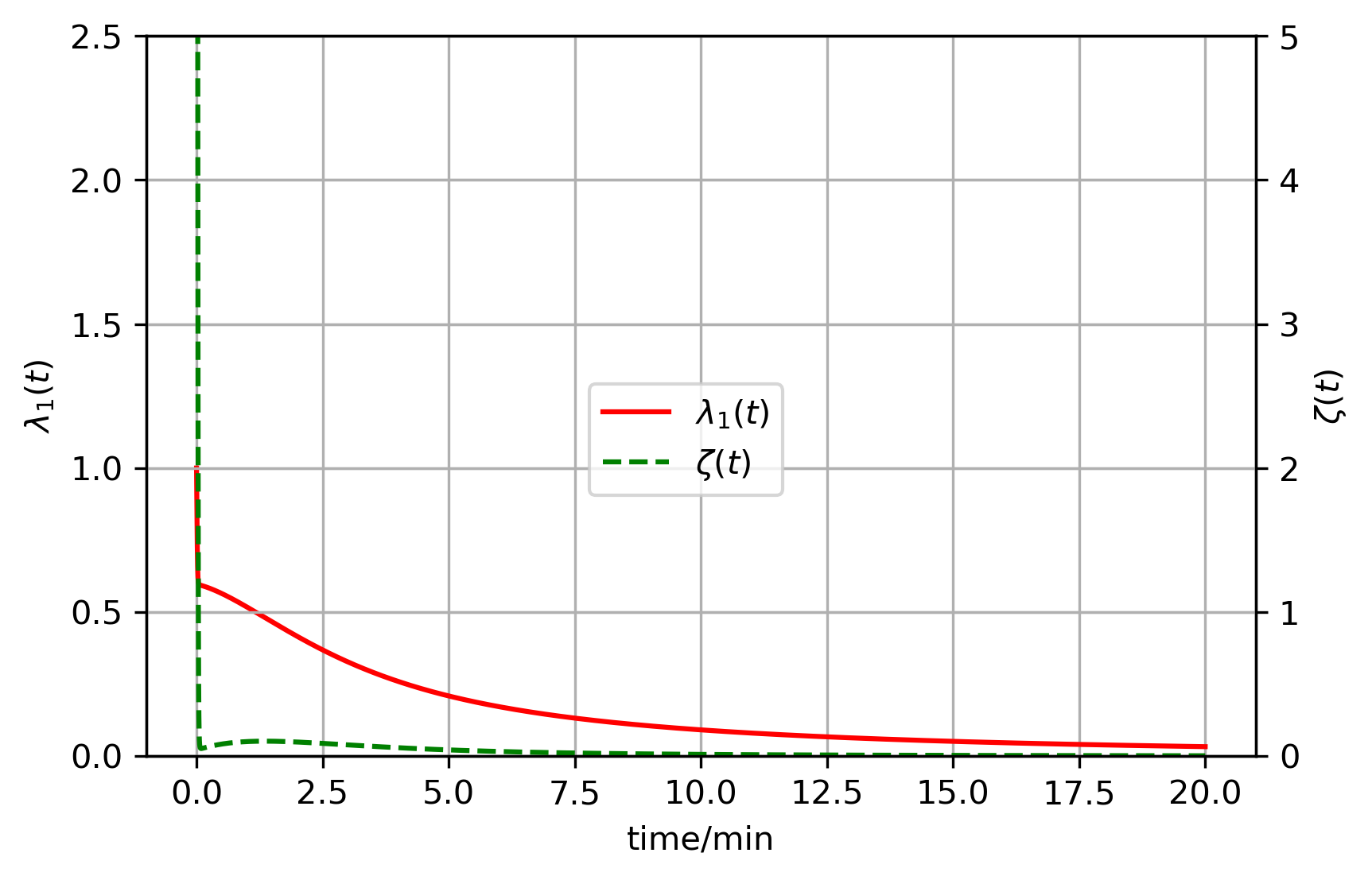}
	\caption{}
	\label{timeseries2}
\end{subfigure}
\caption[]
{\small Numerical verification of stability of the closed-loop system with the vehicle-based UE model}
\label{UE_original_mode1}
\efg

We also show the robustness of the controller. The results in \reff{UE-random-q} and \reff{UE-random-u} are similar to those in \reff{ue}, and the fluctuations are caused by the random demands of HOVs and SOVs. From \reff{UE_cdf_random} and \reff{UE_pdf_random}, we can see that the true VOT distribution can still be accurately estimated even with random demand patterns.

Then, we numerically verify the stability property of the system with constant demands. \reff{phase2} is the phase diagram, which shows the relationship between $\lambda_1(t)$ and $\zeta(t)$. The starting point is (1, 19.9), and the system eventually converges to the ideal state. From \reff{timeseries2}, we can see that $\zeta(t)$ drops near 0 quickly, and then both $\lambda_1(t)$ and $\zeta(t)$ converge to 0 exponentially. With different parameters in the controller and different initial conditions, the simulation results show that the system converges to the ideal state, consistent with the theoretical results in Section 5.2.

\section{Conclusion}

In this paper, we presented a new method to solve the control and estimation problems of a traffic system with HOT lanes.  Different from those in the literature, the method applies even when the lane-choice models and their parameters are unknown to the system operator. By solving the control problem, we determine the dynamic prices, and by solving the estimation problem, we obtain the parameters of the underlying lane-choice models. 

We first defined the traffic flow variables and described the point queue model for traffic flow in Section 2. In Section 3 we presented three different lane-choice models: the multinomial logit model when SOVs share the same value of time, the vehicle-based user equilibrium model when SOVs' values of time are heterogeneous and follow a distribution, and a general lane-choice model. In Section 4, we defined the control and estimation problems and presented a new solution method, by observing that the dynamic price and the excess queueing time on the general purpose lanes are linearly correlated in all the lane-choice models. We also discussed two design principles of dynamic pricing and proved that the principle of maximizing the HOT lanes' throughput is approximately equivalent to the social welfare optimization for logit model. In Section 5, we analytically proved that the equilibrium state of the closed-loop system is ideal, and it is also asymptotically stable. We verified the effectiveness of the solution method with numerical examples in Section 6. 
The new control method can be more readily applicable to managing HOT lanes than existing ones, since it is effective without knowing the underlying lane-choice models or their parameters.

The following are some possible future research topics.
\bi
\item In Section 6, we test the robustness of the dynamic pricing scheme with respect to the random demands. In the follow-up studies, one can test the robustness of the controller by introducing randomness to the underlying traffic flow and lane-choice models, as well as to the measured values. It is also important to examine the robustness with respect to the communication and feedback delay.

\item In this study we focus on the design principles and analytical properties of the pricing scheme and choose not to examine the exact impacts of the parameters $K_1$, $K_2$, $K_3$, and $K_4$. Intuitively, the exact impacts of the parameters depend on the traffic flow and lane-choice models. It will be interesting to develop a formal method for tuning the parameters for the highly nonlinear control system. 
In particular, it is important to test the sensitivity of the closed-loop control system to the coefficients  and determining the optimal coefficients that ensure the stability but minimize the time for the dynamic prices converging to the optimal values. 

\item In this study, the dynamic prices are updated at a high frequency (10 times per second). An interesting topic is to examine the impacts of the updating frequency of the prices. 

\item In Theorem \ref{principle_equivalence} we established the approximate equivalence between the principle of maximizing the HOT lanes' throughput and the social welfare optimization principle. But the proof is based on the assumption of the logit lane-choice model. It will be interesting to examine the equivalence with other lane-choice models. A related topic is to design a pricing scheme that directly optimizes the social welfare.  A related topic is to examine if the pricing scheme designed in this study can also help to maximize the operator's revenues.  

\item Another research topic is to study the control and estimation problems for a freeway corridor with multiple on- and off-ramps and bottlenecks. 
\ei

\section*{Acknowledgments}
The first and second authors would like to thank the ITS-Irvine Mobility Research Program (SB1) for the financial support. We would like to thank the editor and two reviewers for their detailed and helpful comments and suggestions. Discussions with Irene Martinez are greatly appreciated.

\pdfbookmark[1]{References}{references}

\end {document}